# Pressure-induced phonon freezing in the ZnSeS II-VI mixed crystal: phonon-polaritons and *ab initio* calculations

R. Hajj Hussein,[1] O. Pagès,[1,*] A. Polian,[2] A.V. Postnikov,[1] H. Dicko,[1] F. Firszt,[3] K. Strzałkowski,[3] W. Paszkowicz,[4] L. Broch,[1] S. Ravy[5] and P. Fertey[5]

[1] *LCP-A2MC, Institut Jean Barriol, Université de Lorraine, France*

[2] *Institut de Minéralogie, de Physique des Matériaux et de Cosmochimie, Sorbonne Universités – UPMC Université Paris 06, UMR CNRS 7590, F-75005 Paris, France*

[3] *Institute of Physics, Faculty of Physics, Astronomy and Informatics, Nicolaus Copernicus University, Grudziadzka 5/7, 87-100 Toruń, Poland*

[4] *Institute of Physics, Polish Academy of Sciences, 02-668 Warsaw, Poland*

[5] *Synchrotron SOLEIL, L'Orme des Merisiers Saint-Aubin - BP 48 91192 Gif-sur-Yvette Cedex, France*

**Abstract**

Near-forward Raman scattering combined with *ab initio* phonon and bond length calculations is used to study the 'phonon-polariton' transverse optical modes (with mixed electrical-mechanical character) of the II-VI $ZnSe_{1-x}S_x$ mixed crystal under pressure. The goal of the study is to determine the pressure dependence of the poorly-resolved percolation-type Zn-S Raman doublet of the three oscillator [1x(Zn-Se),2x(Zn-S)] $ZnSe_{0.68}S_{0.32}$ mixed crystal, which exhibits a phase transition at approximately the same pressure as its two end compounds (~14 GPa, zincblende→rocksalt), as determined by high-pressure x-ray diffraction. We find that the intensity of the lower Zn-S sub-mode of $ZnSe_{0.68}S_{0.32}$, due to Zn-S bonds vibrating in their own (S-like) environment, decreases under pressure (Raman scattering), whereas its frequency progressively converges onto that of the upper Zn-S sub-mode, due to Zn-S vibrations in the foreign (Se-like) environment (*ab initio* calculations). Ultimately, only the latter sub-mode survives. A similar "phonon freezing" was earlier evidenced with the well-resolved percolation-type Be-Se doublet of $Zn_{1-x}Be_xSe$ [Pradhan *et al*. Phys. Rev. B **81**, 115207 (2010)], that exhibits a large contrast in the pressure-induced structural transitions of its end compounds. We deduce that the above collapse/convergence process is intrinsic to the percolation doublet of a short bond under pressure, at least in a ZnSe-based mixed crystal, and not due to any pressure-induced structural transition.

---

[*] Author to whom correspondence should be addressed:
Email: olivier.pages@univ-lorraine.fr

## I. Introduction

The vibrational properties of the III-V and II-VI $AB_{1-x}C_x$ zinc blende semiconductor mixed crystals have been extensively studied over the past mid-century, both experimentally and theoretically.[1-3] Several models have been developed over this period to describe the lattice dynamics of the mixed crystals. One of the most used was the "modified-random-element-isodisplacement (MREI) model".[4] The MREI model was successful in reproducing the apparent two-mode (also noted 1-bond→1-mode) behavior of several of the crystals studied at that time. Nevertheless, new behaviors were discovered in the last decade, in particular with the $Zn_{1-x}Be_xSe$ zincblende[5] mixed crystal in which one bond, namely the short Be-Se one, was associated with two distinct modes. In order to explain such 1-bond→2-modes behavior, we developed a "percolation model",[6] a simple one operating at one dimension (1D) along the linear chain approximation.[7] Since then we have re-examined the phonon mode behavior of several representative zincblende mixed crystals, including $ZnSe_{1-x}S_x$ studied in this work.[4] Apparently the percolation model applies in all cases.

In this work we explore the pressure dependence of the 1-bond→2-modes percolation doublet by focusing on the particular percolation-type $ZnSe_{1-x}S_x$ zincblende system. A similar study had earlier been done with the leading percolation-type $Zn_{1-x}Be_xSe$ zincblende mixed crystal.[5] The actual extension is interesting in that the two mixed crystals are much different in terms of the pressure-induced structural transitions of their parent compounds, being either similar ($ZnSe_{1-x}S_x$) or highly contrasted ($Zn_{1-x}Be_xSe$). More detail is given at a later stage.

Generally, the percolation scheme is potentially interesting for the study of the pressure-induced structural transitions of mixed crystals. For a better appreciation, we summarize its basic content hereafter. This is further useful to introduce notations.

The percolation scheme is centered on the transverse optical (TO) modes, as detected in a traditional Raman experiment performed in the backscattering geometry. The specificity of the latter modes is that they are purely mechanical in character.[8] As such, they hardly couple, and thus preserve the natural richness of the vibration pattern of such a complex system as a mixed crystal. In contrast, the alternative longitudinal optical (LO) modes couple via their macroscopic electric field reflecting the ionicity of the chemical bonding in a polar crystal, which distorts/obscures the vibration pattern.[6] Technically, a nominally pure-TO Raman insight is achieved in backscattering on a (110)-oriented crystal face, corresponding to allowed TO modes and forbidden LO ones.[9]

Originally the percolation scheme has been developed in order to explain the disconcerting three-modes [1x(Zn-Se),2x(Be-Se)] TO Raman pattern of $Zn_{1-x}Be_xSe$ (see Ref. [5] and Refs. therein), falling beyond the scope of the (at the time admitted) two-modes [1x(A-C),1x(B-C)] scheme for a random $A_{1-x}B_xC$ zincblende mixed crystal, as described by the MREI[4] model. Basically the 1-bond→2-modes percolation-type BeSe-like Raman behavior distinguishes between the Zn- and Be-like nearest-neighbor environments of a Be-Se bond. The elementary oscillators behind the lower and upper BeSe-like Raman modes of $Zn_{1-x}Be_xSe$ can be associated with $Be(Se-Be)Se$ and $Zn(Se-Be)Se$, respectively, in the henceforth used 1D notation. The oscillators are moreover referred to as $TO_{Be-Se}^{Be}$ and $TO_{Be-Se}^{Zn}$ in the following, whereby the subscript and the superscript indicate the considered bond vibration and its host environment, respectively. The corresponding oscillator fractions $f$ in the crystal, which affect the TO Raman intensities, scale as the probabilities of finding one Be atom near another Be atom and near a Zn atom on the (Be,Zn)-substituting 1D-sublattice, i.e. as $f_{Be-Se}^{Be} = x^2$ and as $f_{Be-Se}^{Zn} = x \cdot (1-x)$, respectively, assuming a random Be↔Zn substitution.

When compared with experimental data, the 1-bond→2-modes Raman behavior shows up clearly for the short bond only, corresponding to an effective three-mode Raman behavior in total for a mixed crystal, instead of the nominal four-mode (2 modes per bond) one. This is because the local strain due to the contrast in bond length of the constituent species is mostly accommodated by the short bond in a zincblende mixed crystal, as is well-known from extended-x-ray absorption fine structure (EXAFS) measurements.[10] With this, the variation in the local distortion of a bond, in dependence on whether the latter finds itself either in 'own' or 'foreign' environment, is larger for the short species than for the long one, with concomitant impact on the frequency gap between the corresponding two like sub-modes, being large and negligible, respectively.

Along the same line of reasoning, the percolation-type Raman doublet of a short bond is best resolved in mixed crystals with a large contrast in their bond physical properties. For example, the contrast is large in $Zn_{1-x}Be_xSe$, where the Be-Se bonds are by ~10% shorter[11] and roughly twice more covalent, and hence more rigid,[12] than the Zn-Se ones. Altogether this gives rise to a distinct bimodal BeSe-like Raman signal, characterized by an unusually large frequency gap, of ~40 cm$^{-1}$.[5]

It becomes clear now how the percolation scheme can be used for the study of the pressure-induced structural transitions of mixed crystals. Basically, this is due to the scheme's sensitivity for local environments of a (short) bond. This offers a chance to 'see' how such bonds behave depending on their local ('like' or 'foreign') environment, at, e.g., the approach of a pressure-induced structural transition. We are not aware of any other technique with a similar ability, as was already discussed in Ref. [13].

For example, high-pressure x-ray diffraction, the dedicated technique to study the pressure-induced structural transitions, provides information upon the lattice parameter. This is concerned with the average distance between high-density atomic planes, irrespectively of their constituent species. As such the x-ray diffraction provides a macroscopic-like insight into the lattice relaxation. A more refined insight at the bond scale is obtained via EXAFS measurements, addressing the individual bond species. However, we have checked that even the difference in bond length behind the well-resolved Be-Se Raman doublet of the highly contrasted $Zn_{1-x}Be_xSe$ mixed crystal (see above) is so small (less than 2% from *ab initio* calculations[14]) that it cannot be resolved experimentally by EXAFS.[15,16]

Recently we have performed a pioneering study of the pressure dependence of the BeSe-like percolation doublet of the exemplary $Zn_{1-x}Be_xSe$ mixed crystal (x=0.11, 0.16, 0.24, 0.55), combining Raman scattering in the traditional backscattering geometry and *ab initio* phonon calculations.[5] We observed that the lower sub-mode, which is due to Be-Se bonds vibrating in their own Be-like environment, progressively collapses under pressure, and at the same time converges onto its upper counterpart, due to Be-Se vibrations in the foreign Zn-like environment. Remarkably the exact degeneracy of the two Be-Se sub-modes, coinciding with the full extinction of the lower one, was achieved around 14 GPa, whatever the Be content $x$. It could not be merely fortuitous that the latter pressure approximately corresponds to the rock salt transition of pure ZnSe,[17] thus pointing to a crucial role of the host ZnSe-like matrix behind the collapse/convergence process. We suggested that the latter process reflects, in fact, a progressive 'freezing' of the Be-Se vibrations in their own environment when they are forced to adopt the structure of the host ZnSe-like matrix. As such, the collapse/convergence of the Be-Se vibrations does ultimately stem from large contrast between the pressure-induced structural transitions of the parent compounds ZnSe (at ~14 GPa, Ref. [17]) and BeSe (→NiAs at ~56 GPa, Ref. [18]), and would not be intrinsic to the percolation doublet.

In this work, we further test this idea by performing high-pressure Raman measurements with the alternative percolation-type three-oscillator [1x(Zn-Se),2x(Zn-S)] ZnSe-based $ZnSe_{1-x}S_x$ mixed crystal, whose Raman spectra were recently shown to obey the percolation scheme,[4] with special attention paid to the Zn-S doublet. $ZnSe_{1-x}S_x$ is chosen for the test since it is characterized by nearly identical pressure-induced structural transitions of its parent ZnS and ZnSe compounds (both transit at ~14 GPa, Ref. [19]), probably due to the moderate contrasts in the bond lengths (~4 %) and bond ionicities (~3 %) of its constituent species.[11] The Raman study is completed by *ab initio* phonon and bond length calculations by applying the Siesta code to the prototypal percolation-type S-trio impurity motif (three S atoms in a row) immersed in a large (64-atom) cubic (2x2x2) ZnSe-like supercell. The calculations are repeated at various pressures on both sides of the pressure-induced structural transition of the used $ZnSe_{1-x}S_x$ mixed crystal, as independently determined by high-pressure x-ray diffraction, while maintaining artificially the original structure of the supercell. The *ab initio* insight will thus be independent of the latter transition, *i.e.* intrinsic to the ZnS-like percolation doublet, in contrast with the Raman insight.

High-pressure backscattering Raman measurements at up to ~13 GPa have earlier been performed in this spirit by Basak *et al.*[20] on a polycrystalline $ZnSe_{0.5}S_{0.5}$ sample, supported by calculations of the Γ-like (q=0) one-phonon density of states (abbreviated Γ-projected PhDOS) done within the shell model using disordered $Zn_{32}Se_{16}S_{16}$ 2x2x2 (64-atom) supercells. Owing to the polycrystalline character of the used sample, the reported experimental modes were of the (TO+LO) type. The shell model calculations were accordingly developed within both the TO and LO symmetries, resulting in an overall TO/LO-mixed Γ-projected PhDOS.

Now, due to the large broadening of the two close ZnS-like TO sub-modes at x~0.5 (of about ~20 cm$^{-1}$, comparable to the frequency gap between the two sub-modes), it is difficult from the reported experimental data by Basak *et al.* to decide about any collapse/convergence process within the Zn-S doublet under pressure from ~5 GPa onwards. In fact, the two Zn-S TO sub-modes are hardly distinguishable one from another already at ambient pressure. The shell model calculations seem more rewarding with this respect. While a significant reduction of the Zn-S frequency gap of ~25% is observed when the pressure increases up to 10 GPa, apparently the Zn-S doublet remains finite and well-resolved at any pressure. Besides, no significant collapse is observed. Altogether this seems to oppose to the Be-Se trends in $Zn_{1-x}Be_xSe$, apparently supporting our view that the collapse/convergence processes observed with the Be-Se doublet is primarily due to the large contrast in the pressure-induced structural transitions of the BeSe and ZnSe parent compounds.

However, the TO/LO-mixed character of the experimental/theoretical data reported by Basak *et al.* makes it difficult to decide whether the examined Zn-S doublet by the cited authors is a pure-TO one, in the ideal case, or a LO-contaminated one, which might introduce some bias in the discussion. In fact, Vinogradov *et al.* have shown, based on a careful study by far-infrared reflectivity,[21,22] that the upper Zn-S mode of $ZnSe_{1-x}S_x$ remains quasi TO-LO degenerate throughout the whole composition domain, with a slight but noticeable LO-TO inversion (the LO mode emerging ~5 cm$^{-1}$ – at most – below the TO one).

In order to waive any doubt we search in the present work for a pure-TO Raman insight into the pressure dependence of the Zn-S doublet of $ZnSe_{1-x}S_x$, both experimentally and theoretically. Experimentally, this can be achieved by using a properly oriented (110) monocrystal, as already mentioned. Besides, for an optimal insight we place the analysis at x~0.3 corresponding to quasi intensity matching between the two ZnS-like TO sub-modes.[6] Note that the intensity matching occurs at a lower x value for the Zn-S (x~0.3) doublet than for the Be-Se (x~0.5) one. This is because the Zn-S vibration is sensitive to its local environment at the second-neighbor scale in $ZnSe_{1-x}S_x$, and not at the first-neighbor one as for Be-Se in $ZnBe_xSe_{1-x}$. Details are reported elsewhere.[6]

One major difficulty remains that the resolution of the Zn-S TO Raman doublet of $ZnSe_{1-x}S_x$ is intrinsically poor in the traditional backscattering geometry. In fact, the frequency gap between the two ZnS-like TO sub-modes is even worse at x~0.3 than at x~0.5, hardly reaching ~15 cm$^{-1}$.[6] To circumvent this difficulty, we shift in this work the high-pressure Raman study from the standard backscattering geometry to the unusual near-forward scattering geometry (schematically operating in the 'transmission' mode), searching for those particular TO modes called phonon-polaritons (PP). The reason is that the frequency gap between the two ZnS-like TO sub-modes is magnified in the PP-regime (see Ref. [6], detail is given in the course of the discussion). This is helpful in view to resolve experimentally the pressure dependence of the Zn-S doublet.

Generally, while the phonon-polaritons propagating in the bulk of various compounds have been extensively studied over the past mid-century,[23-29] little attention was awarded to mixed crystals so far. In recent years, the dispersion of the phonon-polaritons propagating in the bulk of several $A_{1-x}B_xC$ mixed crystals has been studied theoretically by Bao and Liang, assuming a crude two-mode [1x(A-C),1x(B-C)] MREI-like TO description.[30,31] On the experimental side we are only aware of the pioneering study of the surface phonon-polariton of the one-mode $Al_{1-x}Ga_xN$ mixed crystal with wurtzite structure done by Ng *et al.*[32] using far-infrared attenuated total reflectance, and of our recent near-forward Raman studies of the percolation-type three-mode $Zn_{1-x}Be_xSe$ (Refs. **33** and **34**) and $ZnSe_{1-x}S_x$ (Refs. **6** and **35**) mixed crystals. As for the high-pressure study of phonon-polaritons, there are no data in the literature, not even for a pure compound.

## II.   Experimental details and *ab initio* method

The used $ZnSe_{0.68}S_{0.32}$ sample was grown from the melt as a large size (~0.8 cm in diameter, ~1 cm in length) (110)-oriented single crystal using the high pressure Bridgman method. Details are given elsewhere (Refs. **36** and **37**). The S content was determined by x-ray diffraction at the laboratory scale assuming a linear dependence of the lattice constant on the composition. From the original ingot two different samples were obtained, one finely ground powder and one thin (~20 µm in thickness) (110)-oriented cleaved-piece with parallel faces, to be used for the high-pressure x-ray diffraction measurements and high-pressure Raman measurements, respectively.

The same membrane diamond anvil cell[38] (DAC) has been used for both experiments with diamonds having a culet of 400 µm in diameter. The powder or cleaved samples were placed, together with ruby balls,[39] into a pressure chamber initially formed by 200-µm-thick rhenium or stainless-steel gaskets preindented to 40 µm and drilled by spark-erosion to 150 µm. The pressure was applied using either neon (x-ray diffraction) or 16:3:1 methanol-ethanol-water mixture (Raman) as a pressure transmitting medium. The pressure was measured via the ruby fluorescence linear scale, with an accuracy of ±0.3 GPa at the maximum pressure.[40,41]

The high-pressure x-ray diffraction measurements were performed at the CRISTAL beamline at SOLEIL synchrotron radiation facility (Gif-sur-Yvette, France) from ambient pressure up to ~24 GPa, using a 0.485 Å x-ray wavelength focused onto a ~40 µm in diameter spot (full width at half maximum) at the sample position. At each pressure, the diffraction pattern was recorded using a plane detector by rotating the DAC within ±10° horizontally while maintaining the center of rotation at a fixed distance of the detector (~33 cm), so as to form large and well-defined portions of concentric diffraction rings on the detector. The resulting two-dimensional image plate data were then turned into intensity versus $2\theta$ plots using the software FIT2D.[42] The peaks fitting and unit cell fitting was carried out using the software DATLAB.[43]

Non polarized high-pressure Raman spectra were taken both in the traditional backward scattering geometry and in the unusual near-forward one at every pressure point. Care was taken that the laser beam was focused onto the same sample area in both cases, from the front and rear of the sample, respectively. Owing to the large optical band gaps of ZnSe (2.7 eV) and ZnS (3.6 eV),[44] $ZnSe_{0.68}S_{0.32}$ is transparent to the used 633.0 nm radiation from a HeNe laser, so that the exact superimposition of the rear and front sample spots could be checked by eye using the visible camera mounted on the microscope of our LabRAM HR Horiba-Jobin Yvon micro-Raman setup. In backscattering the laser beam was focused through a 4x objective with a rather moderate working distance of ~2 cm, leading to a reasonably small laser spot at the sample surface, of approximately 10 µm in diameter. In contrast, the near-forward geometry required to focus the laser beam through a lens with a large working distance (~5 cm), resulting in an enlarged spot at the rear of the sample, roughly double. With the used illumination of merely ~50 mW at the laser exit, the resulting power density at the impact spots was rather moderate, and did not generate any appreciable heating effect. Probably due to the combined effects of the small power density on the sample and of the progressive metallization of the sample when approaching the transition, the Raman signals fainted below the detection level beyond 10 GPa. This fixes the limit for the current high-pressure Raman study, coinciding, in fact, with the hydrostatic limit with the used pressure transmitting medium (see above).[45]

*Ab initio* phonon and bond length calculations were done in the local density approximation by applying the computer code Siesta within the frozen-phonon technique[46] to the ultimate percolation-type S-impurity motif of the $ZnSe_{1-x}S_x$ mixed crystal, namely a pseudo-linear chain of three aligned S atoms (S-trio) immersed in a large (64-atom) cubic (2x2x2) ZnSe-like supercell,[6] after a full relaxation of the individual atom positions and of the lattice constant. The calculations setup was essentially the same as detailed in Ref. 6. Similar calculations were performed at ambient pressure (0 GPa), at a pressure close to the expected transition of the real $ZnSe_{0.68}S_{0.32}$ mixed crystal (10 GPa) and well beyond such transition (20 GPa), while maintaining artificially the original structure in each case. We mention that, in its current version, the Siesta code does not take into account the macroscopic electric field accompanying the polar LO or phonon-polariton vibrations near Γ (q=0). The as-obtained Γ-projected Ph-DOS per atom thus assimilate, in fact, with corresponding Raman signals of purely-mechanical TO modes, as apparent in the conventional Raman spectra taken in the backscattering geometry.

### III. Results and discussion

A useful reference prior to studying the vibrational properties of $ZnSe_{0.68}S_{0.32}$ is the pressure-induced structural transition of this mixed crystal. This is determined by high-pressure x-ray diffraction. The raw ZnSe$_{0.68}$S$_{0.32}$ x-ray diffraction spectra taken at increasing pressure (upstroke) up to ~22 GPa are shown in **Fig. 1a**. Phase transformation, clearly first order, is detected at ~14 GPa. Note a small admixture of the two phases in the reported spectrum at 14.7 GPa in **Fig. 1a.** With no surprise, the considered mixed crystal transforms to rock salt at the same critical pressure as its ZnSe and ZnS parent compounds, within less than 0.5 GPa. The corresponding variation of the lattice parameter versus pressure is reported in **Fig. 1b**.

*1. The reference ZnSe$_{0.68}$S$_{0.32}$ phonon-polariton dispersion (at ambient pressure)*

The theoretical "frequency ($\omega$) vs. $y$" dispersion at ambient pressure of TO modes propagating in the bulk ZnSe$_{0.68}$S$_{0.32}$ mixed crystal is shown in **Fig. 2** (derived from Fig. 10 of Ref. **6**), for reference purpose. $y$ is a dimensionless parameter which conveniently substitutes for the magnitude of the phonon wave vector $q$. It is defined as $\frac{qc}{\omega_1}$, where $\omega_1$ arbitrarily represents the frequency of the TO mode of pure ZnSe (205 cm$^{-1}$) and $c$ is the speed of light in vacuum. The quasi vertical dispersion of the pure transverse electromagnetic wave, namely a photon, at frequencies well beneath ($\omega \to 0$, dotted line) and well beyond ($\omega \to \infty$, dashed line) the phonon resonances are added (thick lines), for the sake of completeness.

Depending on the used scattering geometry, *i.e.* on the used laser excitation and on the scattering angle (s.a.) between the wave vectors of the incident laser beam ($\vec{k}_i$) and of the scattered light ($\vec{k}_s$), one is likely to address different TO regimes. It is all governed by the wave vector conservation rule $\vec{q} = \vec{k}_i - \vec{k}_s$.

In the conventional backscattering geometry (s.a.~180°), $\vec{k}_i$ and $\vec{k}_s$ are (nearly) parallel and opposite, so that $q$ reaches maximum (for a light scattering experiment), of the order of 1% of the Brillouin zone size. This falls far away from the quasi vertical dispersions of a photon at phonon-like frequencies, so that the transverse electric field which is expected to accompany a TO mode in a polar crystal cannot propagate. The TO modes of a polar crystal detected in a traditional backscattering Raman experiment thus consist of purely-mechanical vibrations, somewhat counterintuitively.[5] The corresponding TO frequencies coincide with the reported asymptotic values at large ($q$,$y$) values in **Fig. 2**., *i.e.* three in total [1x(Zn-Se),2x(Zn-S)].

In view to restore the transverse electric field of a TO mode in a polar crystal, one has to get very close to the quasi vertical dispersion of a photon at a phonon-like frequency (note that the concerned $q$ values are much smaller than those achieved in backscattering, by at least two orders of magnitude), hence to use very small s.a. values. As shown in **Fig. 2**, as small s.a. value as 3° is sufficient to retrieve the asymptotic purely-mechanical backscattering-like regime of TO modes. This means that an actual TO mode with mixed mechanical-electrical character, currently referred to as a phonon-polariton, can be detected by Raman scattering only by implementing a near-forward scattering geometry, in which $\vec{k}_i$ and $\vec{k}_s$ are (nearly) parallel and in the same sense (recall the used terminology of a 'transmission' mode in **Sec. I**).

From now on we keep the classical 'TO' notation for the asymptotic purely-mechanical TO modes, being clear that both the latter modes as well as phonon-polaritons basically consist of transverse optical modes.

In **Fig. 2** we are mostly interested in the upper two phonon-polariton branches, which ultimately connect with the frequencies of the purely-mechanical TO modes forming the percolation-type Zn-S doublet at large $(q, y)$ values. These exhibit a characteristic S-like shape governed by two phonon asymptotes, *i.e.* the above-mentioned TO-like one away from Γ and a LO-like one near Γ (being clear that, strictly at Γ, the TO-LO degeneracy occurs).[9] We have already mentioned that the LO mode immediately underneath the upper purely-mechanical Zn-S TO sub-mode is nearly degenerate with the latter mode.[6] It follows that the upper phonon-polariton branch, denoted $PP^+$, is nearly dispersionless. In contrast, the overall S-like distortion is dramatic for the lower branch, denoted $PP^-$. This is because the LO mode immediately underneath the lower purely-mechanical TO sub-mode of the Zn-S bond relates to the alternative Zn-Se bond, which vibrates at a much lower frequency. With this, when $(q, y)$ decreases the frequency gap between the two considered phonon-polariton branches regularly increases from the minimum value of ~15 cm$^{-1}$ in the large $(y, q)$ asymptotic (backscattering-like) regime up to the maximum value of ~60 cm$^{-1}$ right at Γ. Clearly, at least regarding frequencies, the ZnS-like Raman doublet is better resolved in near-forward scattering than in backscattering.

Concerning the Raman intensities, one expects a large Raman efficiency in the Phonon-or-matter-like asymptotic regimes (TO-like away from Γ and LO-like near Γ) and a small one in the intermediary photon-or-light-like regime, approximately corresponding to the inflexion of each phonon-polariton branch. The presumed TO→photon collapse (down to complete extinction) followed by the photon→LO reinforcement were actually modeled and observed experimentally when descending the S-like phonon-polariton dispersion towards Γ, at least for the dominant $PP^-$ feature.[6] Clearly, for an effective insight into the ZnS-like doublet within the phonon-polariton regime, one has to go beyond the collapse regime so as to access the LO-like reinforcement one (otherwise the phonon-polariton just cannot be detected on account of its reduced Raman intensity), which requires to probe ultimately small $q$ values.

Ideally one would like to address exactly the $q=0$ value, corresponding not only to the maximum frequency gap between the two phonon-polariton branches (see above), but also to the maximum Raman intensities of the two modes (assimilating with LO ones then). However, $q=0$ is not achievable experimentally with ZnSe$_{1-x}$S$_x$. This is due to the negative dispersion of its refractive index $n(\omega, x)$ in the visible spectral range. With this, the difference between the frequencies of the incident laser ($\omega_i$) and of the scattered light ($\omega_s$) is increased by the corresponding difference in refractive indexes, at any x value. Therefore, the minimum $q$ value experimentally achievable, given by $|n(\omega_i, x) \times \omega_i - n(\omega_s, x) \times \omega_s|$,[6] remains finite. Now, as the dispersion of the refractive index of $ZnSe_{1-x}S_x$ decreases with frequency, a deeper penetration into the phonon-polariton dispersion is achieved by using lower energy laser lines. The best laser line at hand in our case is the 633.0 nm one, delivered by a HeNe laser.

We have checked in recent work[6] that the 633.0 nm laser line suffices, in fact, to address the LO-like reinforcement regime of the $PP^-$ feature of $ZnSe_{0.68}S_{0.32}$ (while this remained forbidden with the alternative 514.5 nm and 488.0 nm lines from an $Ar^+$ laser). At near-normal incidence (s.a.~0°) of the 633.0 nm laser line the frequency gap between the $PP^-$ and $PP^+$ phonon-polariton branches reaches as much as ~30 cm$^{-1}$, *i.e.* roughly twice that observed in the backscattering geometry. Further the Raman intensity of the lower phonon-polariton $PP^-$ compares with that of the corresponding purely-mechanical TO modes. For a direct insight we show in **Fig. 2** the bimodal ZnS-like Raman signals in both limit geometries, as calculated along the procedure detailed in Ref. **6** using the same set of input parameters, except the s.a. values (specified in **Fig. 2**).

In brief, by using the 633.0 nm laser line in the (nearly) perfect forward scattering geometry (s.a.~0°), all the conditions seem fulfilled, in terms of both frequency gap and Raman intensity, for a reliable Raman study of the ZnS-like doublet of $ZnSe_{0.68}S_{0.32}$ in its pressure dependence.

## 2. *High-pressure near-forward vs. backward Raman studies of $ZnSe_{0.68}S_{0.32}$*

The high-pressure backward (thick curves) and near-forward (thin curves) Raman spectra taken on the cleaved (110)-oriented $ZnSe_{0.68}S_{0.32}$ platelet from ambient pressure up to ~10 GPa are displayed in **Fig. 3**. The two series of Raman spectra are characterized by co-emergence of the TO and LO modes in both the Zn-Se (200 – 250 cm$^{-1}$) and Zn-S (275 – 325 cm$^{-1}$) spectral ranges, though the latter are theoretically forbidden in backward/forward scattering onto (110)-oriented faces of a zinclende crystal (see **Sec. I**). The LO activation is attributed to multi-reflection of the laser beam between the top and rear surfaces of the transparent sample, eventually leading to a (partial) breaking of the wave vector conservation rule governing the Raman scattering.[23]

In spite of the spurious LO features, the TO modes nearby remain clearly visible, both in the Zn-Se and Zn-S spectral ranges. Remarkably, in the Zn-Se spectral range, the TO mode shows up less strongly in the near-forward (F) scattering geometry than in the backward (B) one. This is due to a separation of the two percolation-type Zn-Se sub-modes when entering the phonon-polariton regime (refer to the vertical arrow in **Fig. 3**),[6,35] the two sub-modes in question being otherwise degenerate in B (**Sec. I**). The Zn-S signal is likewise different in the B and F geometries. Whereas in the former (B) geometry, the Zn-S signal reduces to an unique broad band showing up as a compact shoulder on the low-frequency tail of the $LO_{Zn-Se}$ mode, the doublet is clearly visible in the F geometry. This consists, on the one hand, of the nearly dispersionless $PP^+$ mode, possibly contaminated by the spurious backscattering signal due to multi-reflection of the laser beam inside the transparent sample, accompanied, on the other hand, by the $PP^-$ feature, shifted beneath the former $PP^+$ mode by as much as ~30 cm$^{-1}$, as ideally expected when using the 633.0 nm laser line at s.a.~0° (see Fig. 3 of Ref. **35**).

As shown in **Fig. 4** (bottom curves), the theoretical ZnSe$_{0.68}$S$_{0.32}$ $PP^-$ (thin line) and $LO_{Zn-S}$ (dashed-dotted line) Raman lines calculated at normal incidence (s.a.= 0°) for the particular 633.0 nm laser line, using the generic expression for the Raman cross section established in Ref. **6**, coincide in frequency with the corresponding experimental features at ambient pressure within less than 2cm$^{-1}$. No adjustable parameter was used. The important input parameters are the three [1x(Zn-Se),2x(Zn-S)] TO frequencies of ZnSe$_{0.68}$S$_{0.32}$, identified at 210.5, 285.0 and 303.0 cm$^{-1}$ from the backscattering Raman spectrum of the original large-size ingot from which our cleaved sample was taken.[6] We took into account a general feature identified by Vinogradov *et al*.[21,22] based on their careful infrared study of ZnSe$_{1-x}$S$_x$, that, for a satisfactory contour modeling of the Raman and infrared spectra of this mixed crystal throughout the composition domain, the parent-like ZnS oscillator strength should be artificially reduced by ~14% with respect to its nominal value in the pure ZnS crystal.[21] The available Zn-S oscillator strength at 32 at.% S was then evaluated assuming a linear scaling with the Zn-S bond fraction (1-x),[4] and eventually shared in (nearly) equal proportion between the two Zn-S sub-modes (recall the Raman intensity matching between the two backscattering-like TO sub-modes at 30 at.% S).[6] Remaining input parameters from the parent ZnSe and ZnS compounds are available in Ref. **6**.

An interesting question to recollect with the issue raised in this work is whether the frequency gap modifies under pressure, or not? One difficulty is that the $PP^+$ mode does not show up as a distinct feature in our Raman data, owing both to its quasi total extinction at s.a.~0° (see **Fig. 2**) and to its partial screening by the spurious backscattering TO signal. The only reliable markers are the $PP^-$ and $LO_{Zn-S}$ frequencies that remain identifiable from ambient pressure up to 7.25 GPa. The corresponding extreme Raman spectra are compared in **Fig. 4**. The basic trend when the pressure increases is that the ZnSe$_{0.68}$S$_{0.32}$ optical modes harden (see **Fig. 3**), *i.e.* shift to higher frequency, as observed with the pure ZnSe crystal.[47] This is valid for the LO and TO modes and also for the $PP^-$ mode of central interest. Interestingly, the $PP^- - LO_{Zn-S}$ frequency gap decreases between ambient pressure and 7.25 GPa, by as much as ~25%. *A priori* two possible causes may explain such reduction under pressure, either an anomalous convergence of the lower Zn-S TO sub-mode onto the upper one, as earlier observed with the Be-Se doublet of Zn$_{1-x}$Be$_x$Se,[5] or a progressive loss of oscillator strength of the lower Zn-S sub-mode. In the following we explore four possible scenarios combining such causes.

(1)   Equal sharing of the available Zn-S oscillator strength between the two Zn-S TO sub-modes and constant frequency gap between these two at ambient pressure and at 7.25 GPa.

(2) Equal sharing of oscillator strength, and linear convergence of the lower Zn-S TO sub-mode onto the upper one, eventually leading to exact degeneracy at $P_c$~14 GPa corresponding to the ZnSe$_{0.68}$S$_{0.32}$ transition.

(3) Convergence as described in scenario (2), and linear loss of the oscillator strength carried by the lower Zn-S TO sub-mode until complete extinction at $P_c$.

(4) Constant frequency gap and loss of oscillator strength as described in scenarios (1) and (3), respectively.

The criterion for validity is to achieve fair theoretical estimates of simultaneously the $PP^-$ and $LO_{Zn-S}$ frequencies at 7.25 GPa, while using the same set of input parameters.

Technically, we proceed as follows. In their high-pressure Raman study of pure ZnS, Serrano *et al.*[48] found that the pressure-dependencies of the TO and LO frequencies are quasi linear and nearly parallel (within less than 5%) in the considered pressure range (0→7.25 GPa). In a first approximation we assume that the pressure-dependencies of the upper Zn-S TO frequency ($\omega_T$, not observed experimentally) and of the $LO_{Zn-S}$ one ($\omega_L$) are likewise linear and parallel for ZnSe$_{0.68}$S$_{0.32}$. Note that the progressive hardening of the TO and LO modes of the pure ZnS crystal, of ~5 cm$^{-1}$/GPa,[48] suffices *per se* to reduce significantly the parent-like ZnS oscillator strength to be used for ZnSe$_{0.68}$S$_{0.32}$ at 7.3 GPa, as compared with the reference value at ambient pressure. As the $\varepsilon_\infty$ values of pure ZnS and pure ZnSe remain quasi stable under pressure, at least up to 20 GPa,[49] and thus presumably also the Zn$_{0.68}$S$_{0.32}$ one, the reduction in the ZnS oscillator strength, which expresses as $\varepsilon_\infty \cdot (\omega_L^2 - \omega_T^2)/\omega_T^2$,[4] amounts to ~25%. We recall that the parent ZnS oscillator has to be further reduced by 14% before being used with Zn$_{0.68}$S$_{0.32}$. Last, in a crude approximation, we neglected the pressure dependence of the ZnSe$_{0.68}$S$_{0.32}$ refractive index. However, as detailed in the **Appendix section**, this has little importance for our use.

The as-obtained theoretical $PP^-$ Raman lineshapes calculated along scenarios (1)-to-(4) are superimposed onto the experimental Raman spectrum at 7.25 GPa in **Fig. 4**, for comparison. In each case the individual Zn-S TO frequencies were slightly adjusted so as to achieve a perfect matching between the theoretical and experimental $LO_{Zn-S}$ frequencies. While scenarios (1) and (2) fail to reproduce the experimental $PP^-$ frequency, scenarios (3) and (4) are rather successful in this respect. The Raman data are thus not conclusive regarding whether the ($TO_{Zn-S}^S \rightarrow TO_{Zn-S}^{Se}$) converge or not under pressure, but point towards a loss of oscillator strength of the lower Zn-S TO sub-mode.

### 3. Ab initio phonon / bond length calculations at the S-dilute limit

In view to solve the pending issue concerning the convergence, or not, of the lower Zn-S sub-mode onto the upper one under pressure, we resort to *ab initio* phonon and bond length calculations. The same calculation pattern (the structure relaxation under a target pressure followed by frozen-phonon calculation), using the prototype percolation-type S-impurity motif (a trio of pseudo-aligned S atoms at closest anion sites – abbreviated S-trio) was repeated at ambient pressure, 10 and 20 GPa, while maintaining the original structure at any pressure. The as-obtained distributions of Zn-Se (top inset, host matrix) and Zn-S (bottom inset, S-trio motif) bond lengths, and the corresponding TO-like (see **Sec. II**) $\Gamma$−projected Ph-DOS per Se (thin curves) and S (thick curves) atoms, are displayed in **Figs. 5** and **6**, respectively. First, we discuss the basic trends at ambient pressure, for reference purpose.

As expected (see **Sec. I**), the long Zn-Se bonds have quasi identical lengths whether located close to the S-trio or away from it. In contrast the short Zn-S bonds are long along the S-trio chain, i.e. when they are self-connected, and short perpendicularly to the S-trio chain, i.e. when they connect to the ZnSe-like host medium – as schematically represented in **Fig. 5**. A similar pattern was earlier derived for the prototypic percolation-type Be doping of ZnBeSe.[14]

Based on the "rule of thumb" that shorter bonds vibrate at a higher frequency, one may well transpose the above picture for the lattice relaxation of the S-trio (structural aspect) to its lattice dynamics (vibrational aspect). This works rather well, in fact. Indeed while the $\Gamma$−projected Ph-DOS per Se atom consists of a unique feature, its S-counterpart is clearly bimodal. More precisely, out of the nine possible atom vibration patterns produced by the S-trio (three degrees of freedom per atom), the Zn-S stretching along the pseudo-linear S-trio chain involving the *longer* Zn-S bonds (see above), emerges at significantly lower frequency (~300 cm$^{-1}$) than the remaining modes of the series (regrouped around 315 cm$^{-1}$ into the so-called trio-bending band). These latter include individual vibration patterns shown in Fig. 5 of Ref. **6**, numbered 184 to 192 therein, in order of increasing frequency among the 64×3 modes of the supercell. The bending modes of the S-trio involve the stretching of the *shorter* Zn-S bonds from the side of the S-trio in the "foreign" Se-like environment. Altogether, this is consistent with the lower (trio-stretching mode) and upper (trio-bending band) $\Gamma$−projected PhDOS of the S-trio being described at 1D within the percolation scheme in terms of the stretching of Zn-S bonds within their own S-like environment and within the foreign Se-like one, respectively (see **Sec. I**).

Under pressure, the general trend is that the Zn-Se and Zn-S distributions of bond lengths sharpen and re-center towards smaller values. The unimodal-like Zn-Se distribution is illustrative with both respects, but the Zn-S distribution is most impacted since it changes from bimodal (ambient pressure) to unimodal (20 GPa).

Once again, the lattice relaxation transposes fairly well to the lattice dynamics (refer to the above "rule of thumb"). The analogy is straightforward for Zn-Se, for which a unique $\Gamma$−projected PhDOS feature is anyway observed, now shifted to higher energy with respect to ambient pressure. For Zn-S the situation is less clear since a multi-mode vibration survives at 20 GPa (not surprisingly, since the lattice dynamics, that depends also on the distribution in bond angles, is intrinsically more complicated than the bond length distribution). Now, the crucial feature for our concern is that the frequency gap between the unique trio-stretching mode and the Si-trio bending band drastically reduces under pressure. While the gap is well-resolved at ambient pressure (299 – 307 cm$^{-1}$), it has become narrow at 10 GPa (356 – 359 cm$^{-1}$) and has just disappeared at 20 GPa (the trio-stretching mode then emerges right in the middle, i.e. at ~418 cm$^{-1}$, of the trio-bending band, covering the 407 – 430 cm$^{-1}$ spectral domain). At this limit, we may safely state that the original Zn-S doublet (ambient pressure) has coalesced into a proper singlet (20 GPa).

Now, we examine in greater detail what is the microscopic mechanism at the origin of the pressure-induced doublet→singlet ZnS-like coalescence. At the simplest level, one may well feel satisfied with the above mapping of the lattice relaxation onto the lattice dynamics. However, be it the only cause, the individual vibration patterns behind each ZnS-like Γ−projected Ph-DOS would remain the same (or, at least, preserved within some relevant combination) as the pressure changes. This is actually so for the individual trio-bending modes, as already mentioned, and also for the remaining trio-stretching one up to 10 GPa. As ideally expected for a Γ-like vibration, the atom displacement behind the latter mode consists of the three S atoms vibrating in phase and with comparable magnitude along the chain (top of **Fig. 6**). All Zn-S bonds forming the S-trio motif are involved, as schematically indicated beneath the actual vibration pattern. Now, the latter vibration pattern changes at 20 GPa, suggesting a more refined mechanism of coupling between the lattice relaxation and dynamics. Basically the vibration of the central S atom along the S-trio chain, in its own S-like environment, becomes hindered at 20 GPa (bottom of **Fig. 6**). Only the side Zn-S bond of the S-trio stretch at this limit, as schematically emphasized beneath the reported vibration pattern. Interestingly, a similar mechanism of 'phonon-freezing' was earlier put forward to explain at the microscopic scale the pressure-induced collapse of the percolation-type Be-Se Raman doublet of Zn$_{1-x}$Be$_x$Se into a singlet, also based on *ab initio* calculations (using the prototypal percolation-type Be-duo motif then).

Retrospectively, we note that such 'freezing' of the lower Zn-S sub-mode under pressure, as observed *ab initio*, is consistent with the retained scenario, at the term of the high-pressure ZnSe$_{0.68}$S$_{0.32}$ Raman study (see **Sec. III-2**), of a pressure-induced loss of oscillator strength for that sub-mode.

Summarizing, the pressure-induced 'phonon-freezing' is observed for both the percolation doublets of the short bonds of the ZnSe$_{1-x}$S$_x$ (this work) and Zn$_{1-x}$Be$_x$Se (Ref. **5**) mixed crystals, characterized by nearly identical and highly-contrasted structural phase transitions of their parent compounds, respectively. Moreover, in the particular case of ZnSe$_{1-x}$S$_x$, the 'phonon-freezing' was identified *ab initio* by 'forcing' the zincblende structure (20 GPa) well beyond the actual pressure-induced zincblende→rocksalt transition of this mixed crystal (occurring at ~14 GPa, see **Sec. III**). Altogether, this indicates that the 'phonon-freezing' is intrinsic to the percolation doublet of the short bond, at least in a ZnSe-based mixed crystal, and not related to its pressure-induced structural transition.

## IV. Conclusion

We combine Raman scattering and *ab initio* calculations to test an idea earlier formulated at the occasion of the high-pressure Raman study of the zincblende $Zn_{1-x}Be_xSe$ mixed crystal[5] that the observed pressure-induced convergence of the percolation-type doublet due to its short Be-Se bond basically relates to the unusually large contrast in the Zn-Se and Be-Se bond physical properties, eventually leading, at the macroscopic scale, to highly-contrasted pressure-induced structural transitions of its parent compounds. A choice system for the test is the zincblende $ZnSe_{1-x}S_x$ mixed crystal. The latter exhibits a distinct percolation doublet of its short Zn-S bond,[6] albeit worse resolved than the Be-Se one of $Zn_{1-x}Be_xSe$, and, moreover, the pressure-induced structural transitions of its parent compounds are nearly identical. For more clarity we place the analysis around the sensitive composition of 33 at.% S, corresponding to quasi intensity matching between the two Zn-S Raman sub-modes forming the Zn-S doublet, and perform the high-pressure Raman study in the phonon-polariton regime using an unusual near-forward scattering geometry.

The pressure-dependence of the Raman frequency of the dominant phonon-polariton, related to the lower Zn-S sub-mode – due to Zn-S vibrations in their own S-like environment, indicates a progressive loss of oscillator strength carried by the latter mode under pressure. On the other hand the *ab initio* calculations reveal a progressive convergence of the lower Zn-S sub-mode onto the upper one under pressure, leading to degeneracy at 20 GPa. This is due to a pressure-induced 'freezing' of the Zn-S bonds when they stretch in their own S-like environment, possibly explaining the above mentioned loss of oscillator strength evidenced at high pressure by Raman scattering.

As a similar 'freezing' was earlier evidenced for the short Be-Se bonds of $Zn_{1-x}Be_xSe$,[5] we deduce that the pressure-induced convergence and loss of oscillator strength are intrinsic to the percolation-type vibrational doublet of a short bond, at least in a ZnSe-based mixed crystal, and not due to the contrast in the pressure-induced structural transitions of the parent compounds. The physical reason why the short bonds vibrating in their own environment 'freeze' under pressure remains unclear at present.

**Acknowledgements**
We would like to thank P. Franchetti for technical assistance in the Raman measurements. This work was supported by the European funding of Region Lorraine under project FEDER-Percalloy n°. presage 34619.

**Appendix Section**

In this section we investigate to which extent the predictions concerning the pressure-induced shift of the $PP^-$ feature of $ZnSe_{0.68}S_{0.32}$ reported in **Sec. III-2** are modified by taking into account the pressure-dependence of the refractive index.

In their *ab initio* study of the pure ZnS and ZnSe compounds, Khenata *et al.*[49] did predict an overall shift to higher energies of all optical energy transitions under pressure, including the fundamental absorption edge (optical band gap), which basically determines the dispersion of the refractive index near the used laser radiation (633.0 nm in our case). Under an increase of the ambient pressure to 15 GPa, the overall shift amounts to ~0.6 eV for ZnS, roughly twice that predicted for ZnSe. If we linearly interpolate for $ZnSe_{0.68}S_{0.32}$ at ~7 GPa, the corresponding shift for its optical band gap amounts to ~0.2 eV. Considering further that the optical band gap of $ZnSe_{1-x}S_x$ varies (quasi) linearly with the composition $x$, as was recently demonstrated by Zafar *et al.*[50] in their *ab initio* study done for a series of (ordered) mixed crystals, it follows that the $Zn_{0.68}S_{0.32}$ optical band gap enlarges from ~2.96 eV at ambient pressure (300 K) to ~3.16 eV at 7.25 GPa. In a rough approximation this results in an overall shift of the wavenumber dependence of the refractive index of $ZnSe_{0.68}S_{0.32}$ measured at ambient pressure by ~1613 cm$^{-1}$ towards higher wavenumber. Besides, we recall that the (phonon-like) $\varepsilon_\infty$ dielectric constants of pure ZnSe and pure ZnS, which fix an asymptotic limit for the corresponding refractive indexes, remain constant with pressure.[49] Therefore we expect the same for the related $ZnSe_{0.68}S_{0.32}$ mixed crystal. The pressure-induced shift of the wavenumber dependence of the refractive index along the abscissa axis is schematically indicated in **Fig. A1a**.

Such shift does not dramatically change the local dispersion of refractive index around the used 633.0 nm laser line. Indeed, the difference in refractive indexes between, e.g., the incident laser beam and the scattered light at the $PP^-$ frequency remains nearly unaffected by the use of either original or shifted curves, as apparent in **Fig. A1a**. For this reason the shift has negligible impact on the $PP^-$ frequency calculated at normal incidence (s.a.= 0°) for the 633.0 nm laser line. Only a moderate softening (and strengthening) is predicted, of at most 2 cm$^{-1}$, as shown in **Fig. A1b**. This is not challenging for the conclusions drawn at the term of **Sec. III-2**.

**References**

[1] R. J. Elliott, J.A. Krumhansl and P.L. Leath, Rev. Mod. Phys. **46** (1974) 465.

[2] D. W. Taylor, in Optical Properties of Mixed Crystals, edited by R. J. Elliott and I.P. Ipatova (North-Holland, Amsterdam, 1988), chap. 2, p. 35.

[3] S. Adachi, in Properties of Semiconductor ns: Group-IV, III-V and II-VI Semiconductors, edited by John Wiley & Sons (Great Britain, Chippenham, 2009), chap. 4, p. 99.

[4] I. F. Chang and S. S. Mitra, Adv. Phys. 20 (1971) 359.

[5] G. K. Pradhan, C. Narayana, O. Pagès, A. Breidi, J. Souhabi, A.V. Postnikov, S.K. Deb, F. Firszt, W. Paszkowicz, A. Shukla and F. El Hajj Hassan, Phys. Rev. B **81** (2010) 115207.

[6] R. Hajj Hussein, O. Pagès, S. Doyen-Schuler, H. Dicko, A.V. Postnikov, F. Firszt, A. Marasek, W. Paszkowicz, A. Maillard, L. Broch and O. Gorochov, J. Alloys and Compounds **644** (2015) 704.

[7] Due to the quasi vertical dispersion of the light, Raman scattering probes the neighborhood of Γ ($q=0$). At this limit the information on the actual position $\vec{r}$ of an atom in the crystal, as apparent in the propagation term of a plane wave ideally describing a lattice vibration, disappears. Therefore one may as well use a scalar (uni-dimensional, 1D) description of the crystal lattice (along the linear chain approximation) instead of the real vectorial (three-dimensional, 3D) one.

[8] M. Born and K. Huang, Dynamical Theory of Crystal Lattices, (Clarendon, Oxford University Press, 1954).

[9] P. Y. Yu and M. Cardona, in Fundamentals of Semiconductors, edited by Springer-Verlag (Germany, Berlin, 2010), chap. 7, p. 345.

[10] A. Balzarotti, N. Motta, A. Kisiel, M. Zimnal-Starnawska, M. T. Cyzyk and M. Podgorny, Phys. Rev. B **31**, 7526 (1985).

[11] N.E. Christensen, S. Satpathy and Z. Pawlowska, Phys. Rev. B **36** (1987) 1032.

[12] R.M. Martin, Phys. Rev. B **1** (1970) 4005.

[13] O. Pagès, R. Hajj Hussein and V.J.B. Torres, J. Appl. Phys. 114 (2013) 033513.

[14] A. V. Postnikov, O. Pagès and J. Hugel, Phys. Rev. B 71 (2005) 115206.

[15] T. Ganguli, J. Mazher, A. Polian, S. K. Deb, F. Villain, O. Pagès, W. Paszkowicz and F. Firszt, J. Appl. Phys. **108** (2010) 083539.

[16] The ability of the percolation scheme to distinguish between different environments of like bonds in a mixed crystal was, in another context of the diamond-type $Si_{1-x}Ge_x$ mixed crystal,[13] recently used to formalize an intrinsic ability behind Raman scattering to probe the nature of the substitution disorder at the local scale, as to whether the atom substitution is random or not (*i.e.* due to some clustering/anticlustering). At this occasion a formal 'Raman parameter' $\kappa$ of local ordering has been introduced within a generalized $\kappa-$dependent version of the percolation scheme, to measure any deviation with respect to the ideal case of a random substitution.

[17] M. I. McMahon and R. J. Nelmes, Phys. Status Solidi B **198** (1996) 389.

[18] H. Luo, K. Ghandehari, R.G. Greene, A. L. Ruoff, S. S. Trail and F. J. DiSalvo, Phys. Rev. B **52** (1995) 7058.

[19] A. Mujica, A. Rubio, A. Muñoz and R. J. Needs, Rev. Mod. Phys. **75** (2003) 863.

[20] T. Basak, M. N. Rao, S. L. Chaplot, N. Salke, R. Rao, R. Dhanasekaran, A. K. Rajarajan, S. Rols, R. Mittal, V. B. Jayakrishnan and P. U. Sastry, Physica B **433** (2014) 149.

[21] E. A. Vinogradov, B. N. Mavrin, N. N. Novikova, and V. A. Yakovlev, Phys. Sol. State 48 (2006) 1940.

[22] E. A. Vinogradov, B. N. Mavrin, N. N. Novikova, V. A. Yakovlev, and D. M. Popova, Laser Phys. **19** (2009) 162.

[23] C. H. Henry and J. J. Hopfield, Phys. Rev. Lett. **15** (1965) 964.

[24] S. P. S. Porto, B. Tell and T. C. Damen, Phys. Rev. Lett. **16** (1966) 450.

[25] N. Marschall and B. Fischer, Phys. Rev. Lett. **28** (1972) 811.

[26] D. J. Evans, S. Ushioda and J. McMullen, Phys. Rev. Lett. **31** (1973) 369.

[27] D. L. Mills and A. A. Maradudin, Phys. Rev. Lett. **31** (1973) 372.

[28] D. L. Mills and E. Burstein, Rep. Prog. Phys. **37** (1974) 817.

[29] J.-I. Watanabe, K. Uchinokura and T. Sekine, Phys. Rev. B **40** (1989) 7860.

[30] J. Bao and X. X. Liang, J. Phys.: Condens. Matter **18** (2006) 8229.

[31] J. Bao and X. X. Liang, J. Appl. Phys. **104** (2008) 33545.

[32] S. S. Ng, Z. Hassan and H. A. Hassan, Appl. Phys. Lett. **91** (2007) 081909.

[33] R. Hajj Hussein, O. Pagès, F. Firszt, W. Paszkowicz and A. Maillard, Appl. Phys. Lett. 103 (2013) 071912.

[34] H. Dicko, O. Pagès, R. Hajj Hussein, G. K. Pradhan, C. Narayana, F. Firszt, A. Marasek, W. Paszkowicz, A. Maillard, C. Jobard, L. Broch and F. El Hajj Hassan, J. Raman Spectrosc., published online, DOI: 10.1002/jrs.4817.

[35] R. Hajj Hussein, O. Pagès, F. Firszt, A. Marasek, W. Paszkowicz, A. Maillard and L. Broch, J. Appl. Phys. **116** (2014) 083511.

[36] F. Firszt, S. Łęgowski, H. Męczyńska, J. Szatkowski, W. Paszkowicz, K. Godwod, J. Domagała, M. Kozielski, M. Szybowicz, M. Marczak, Proc. 2[nd] Int. Symp. on Blue Lasers and Light Emitting Diodes, Chiba, Japan, Sept 29–Oct. 2, (1998) p.335-338.

[37] F. Firszt, A. Wronkowska, A. Wronkowski, A.S. Łęgowski, A. Marasek, H. Męczyńska, M. Pawlak, W. Paszkowicz, J. Zakrzewski, K. Strzałkowski, Cryst. Res. Technol. **40** (2005) 386.

[38] J. C. Chervin, B. Canny, J. M. Besson, and P. Pruzan, Rev. Sci. Instrum. **66** (1995) 2595.

[39] J. C. Chervin, B. Canny, and M. Mancinelli, High Press. Res. **21** (2001) 305.

[40] H. K. Mao, J. Xu, P. M. Bell, J. Geophys. Res. **91** (1986) 4673.

[41] C.-S. Zha, H. K. Mao, and R. J. Hemley, Proc. Natl. Acad. Sci. U.S.A. **97** ( 2000) 13494.

[42] A. P. Hammersley, ESRF Internal Report N°. ESRF97HA02T, 1997 (unpublished); A. P. Hammersley, S. O. Svensson, M. Hanfland, A. N. Fitch and D. Häusermann, High. Press. Res. **14** (1996) 235.

[43] Kindly provided by K. Syassen, Max Planck Institut für Festkörperphysik, Stuttgart, Germany.

[44] S. S. Devlin, in Physics and Chemistry of II-VI Compounds, edited by M. Aven and J. S. Prener (North-Holland Publishing Company, Amsterdam, 1967), p. 603.

[45] S. Klotz, J. C. Chervin, P. Munsch, G. Le Marchand, J. Phys. D: Appl. Phys. **42** (2009) 075413.

[46] J. M. Soler, E. Artacho, J. D. Gale, A. García, J. Junquera, P. Ordejón and D. Sánchez-Portal, J. Phys.: Condens. Matter 14 (2002) 2745.

[47] B. A. Weinstein, Solid State Commun. **24** (1977) 595.

[48] J. Serrano, A. Cantarero, M. Cardona, N. Garro, R. Lauck, R. E. Tallman, T. M. Ritter and B. A. Weinstein, Phys. Rev. B **69** (2004) 014301.

[49] R. Khenata, A. Bouhemadou, M. Sahnoun, A. H. Reshak, H. Baltache and M. Rabah, Comput. Mat. Sci. **38** (2006) 29.

[50] M. Zafar, S. Ahmed, M. Shakil, M. A. Choudhary and K. Mahmood, Chin. Phys. B **24** (2015) 076106.

# Figure captions

**Fig. 1:** (a) High-pressure x-ray diffraction curves taken at increasing pressure with the ZnSe$_{0.68}$S$_{0.32}$ mixed crystal; and (b) corresponding pressure-dependence of the lattice parameter in the zincblende (ZB) and rocksalt (RS) structures. In part (a) the relevant families of atomic planes giving rise to a particular X-ray line are indicated. The stars indicate diffraction lines from the gasket.

**Fig. 2:** Theoretical dispersion of the ZnSe$_{0.68}$S$_{0.32}$ phonon-polaritons (thick curves) at ambient pressure. The accessible part of the phonon-polariton regime when using the 633.0 nm laser line is delimited by two oblique dispersions curves (dashed-dotted) given by the wave vector conservation rule $\vec{q} = \vec{k}_i - \vec{k}_s$ governing the Raman scattering. The corresponding bimodal ZnS-like Raman signals in the nearly perfect forward geometry (s.a.~0.3°) and assimilating with that obtained in the standard backscattering geometry (s.a.~180°) are shown.

**Fig. 3:** High-pressure ZnSe$_{0.68}$S$_{0.32}$ Raman spectra taken in the backward (thick curves) and near-forward (thin curves, s.a.~0°) scattering geometries using the 633.0 nm laser line.

**Fig. 4:** Combined contour modeling of the ZnS-like PP (plain line) and LO (dashed-dotted line) ZnSe$_{0.68}$S$_{0.32}$ near-forward Raman signals (s.a.~0°) taken at ambient pressure (bottom spectrum) and at ~7 GPa (top spectrum). In the latter case, different scenarios (1-to-4) are considered depending on the pressure dependencies of the oscillator strength (o.s.) and frequency of the lower Zn-S sub-mode, from which the dominant $PP^-$ feature proceeds.

**Fig. 5:** *Ab initio* calculated Zn-Se (top) and Zn-S (bottom) distributions of bond lengths in their pressure dependence for the prototypal percolation-type S-trio impurity motif immersed in a large 64-atom ZnSe-like supercell with zincblende structure. The particular Zn-S bonds of the S-trio involved in the bimodal Zn-S pattern are schematically indicated at ambient pressure.

**Fig. 6:** Pressure-dependent Γ-projected PhDOS per Se (clear curves) and S (dark curves) atoms, assimilating with the Raman signals of corresponding purely-mechanical TO modes, calculated *ab initio* by using the prototypal percolation-type S-trio impurity motif immersed in a large 64-atom ZnSe-like supercell with zincblende structure. The vibration pattern behind the lower Zn-S sub-mode, due to Zn-S stretching in its own S-like environment, is indicated at each pressure. The concerned Zn-S bonds are schematically emphasized underneath.

**Fig. A1:** Assumed variation of the ZnSe$_{0.68}$S$_{0.32}$ refractive index between ambient pressure (filled symbols) and 7 GPa (open symbols), corresponding to an overall shift towards higher energy (blue shift) by the same amount as the fundamental absorption edge (a), and theoretical ZnS-like $PP^-$ Raman lineshapes calculated at 7 GPa while taking into account (1', filled symbols) or not (1, open symbols, see **Fig. 4**) such variations (b). In part (a) the used values of the refractive index for the incident laser and for the scattered light at, e.g., the $PP^-$ frequency ($\omega_S^{PP}$), are indicated in each case, in reference to the raw Raman spectrum taken at 7 GPa. The frequency of the incident laser ($\omega_i$) is indicated for reference purpose.

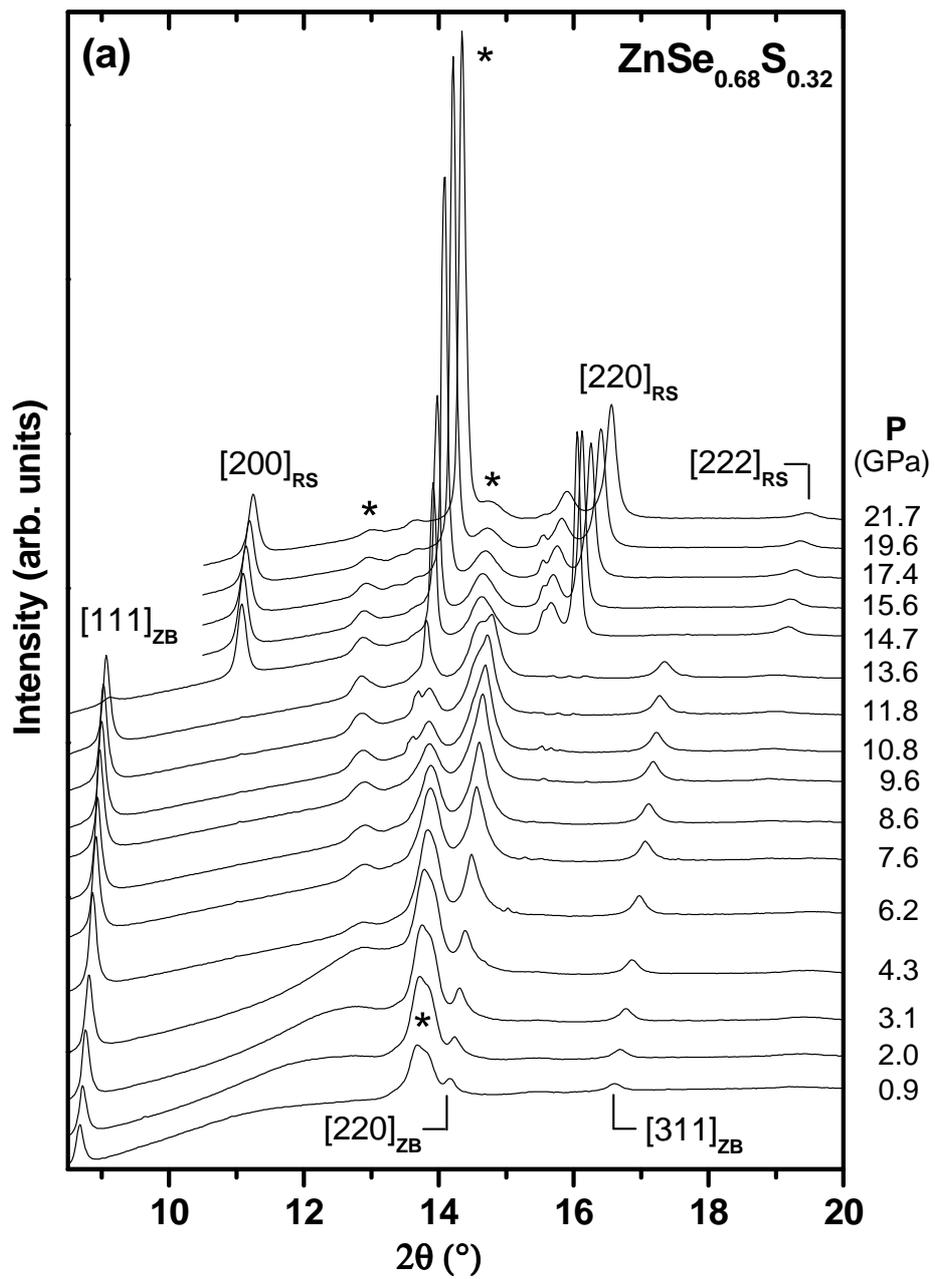

**Figure 1a**

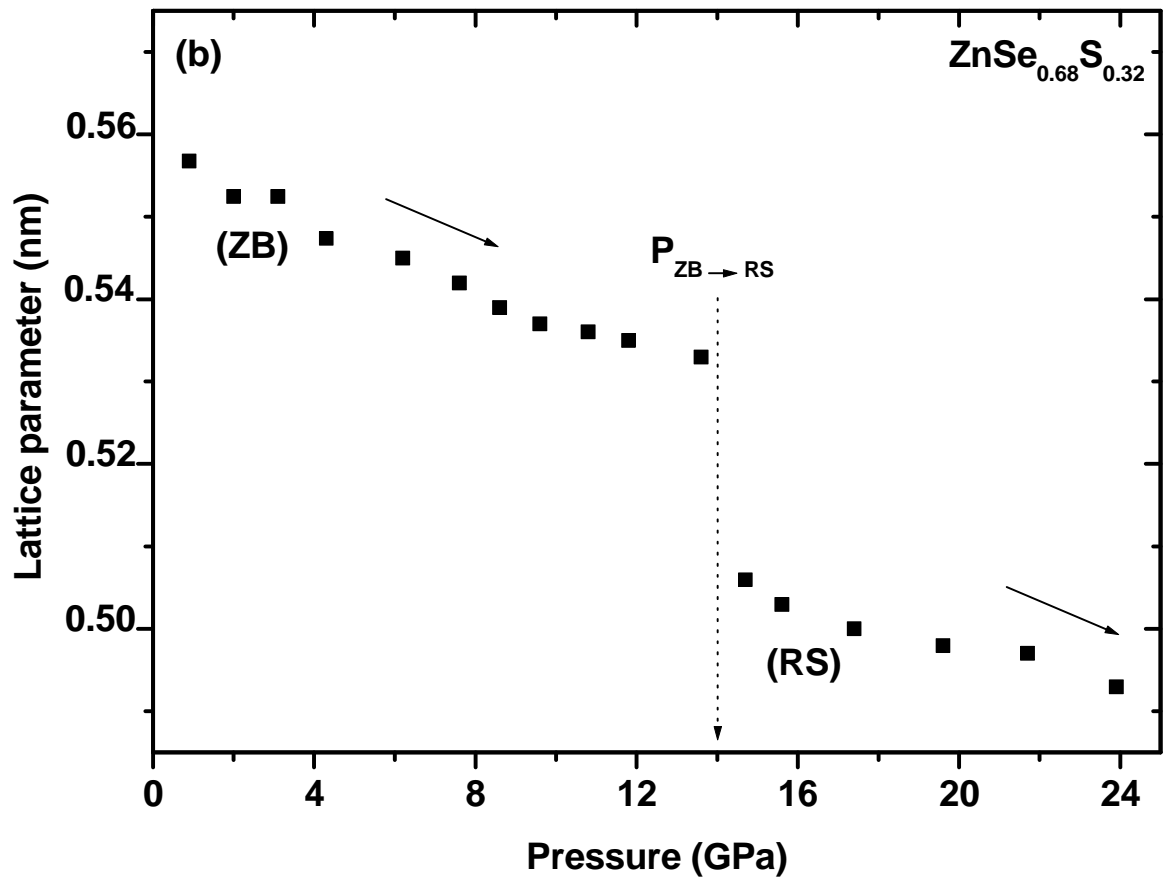

Figure 1b

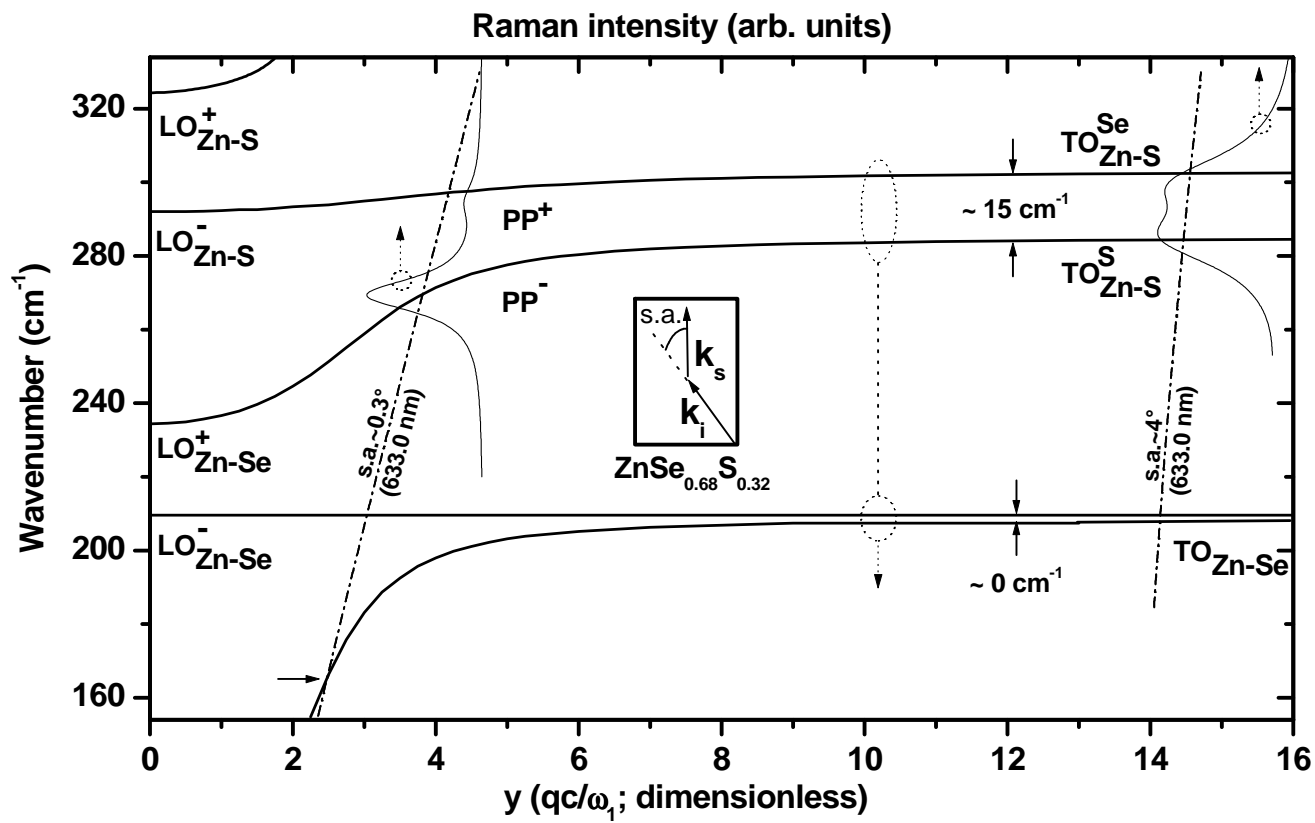

**Figure 2**

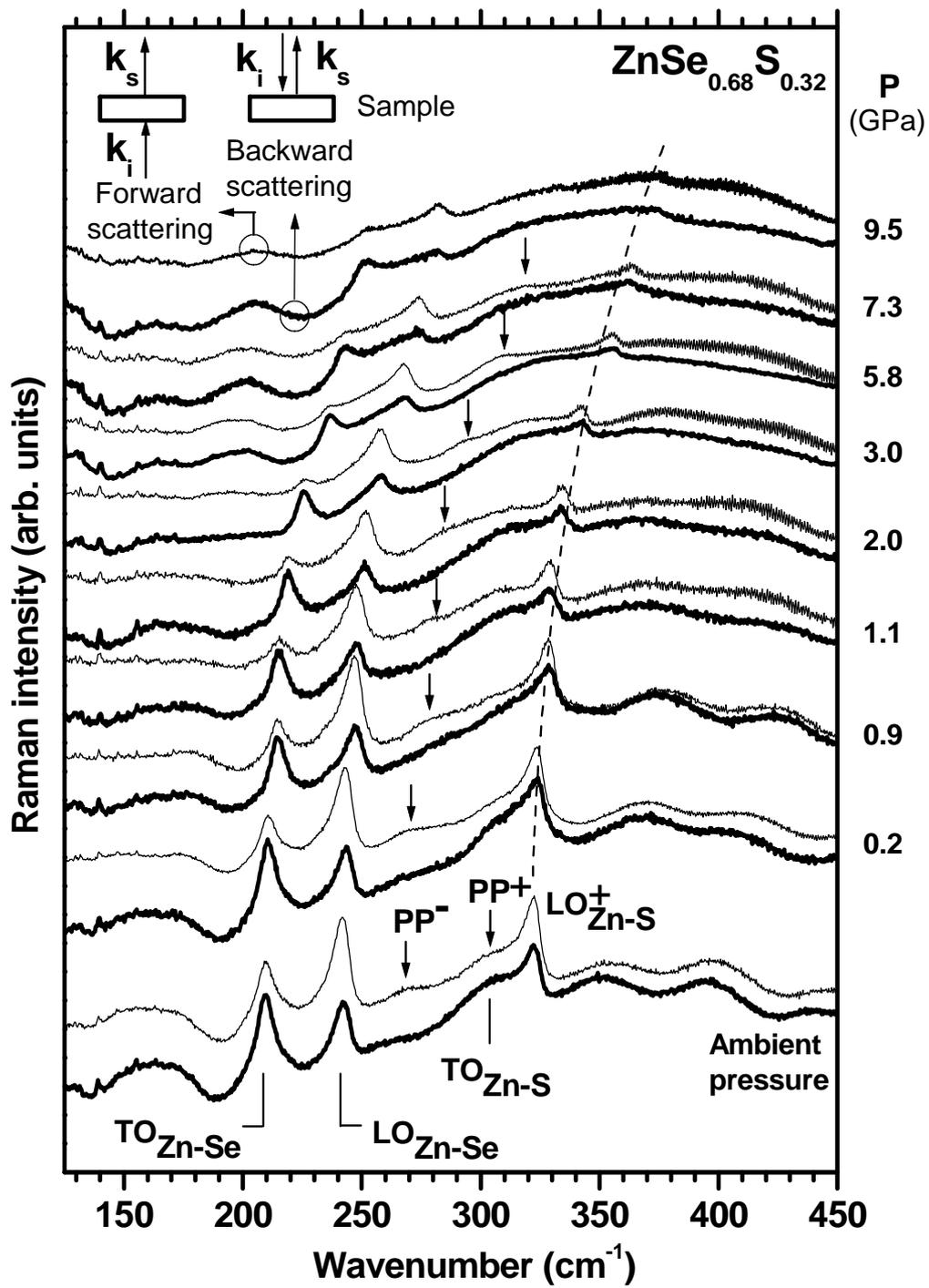

Figure 3

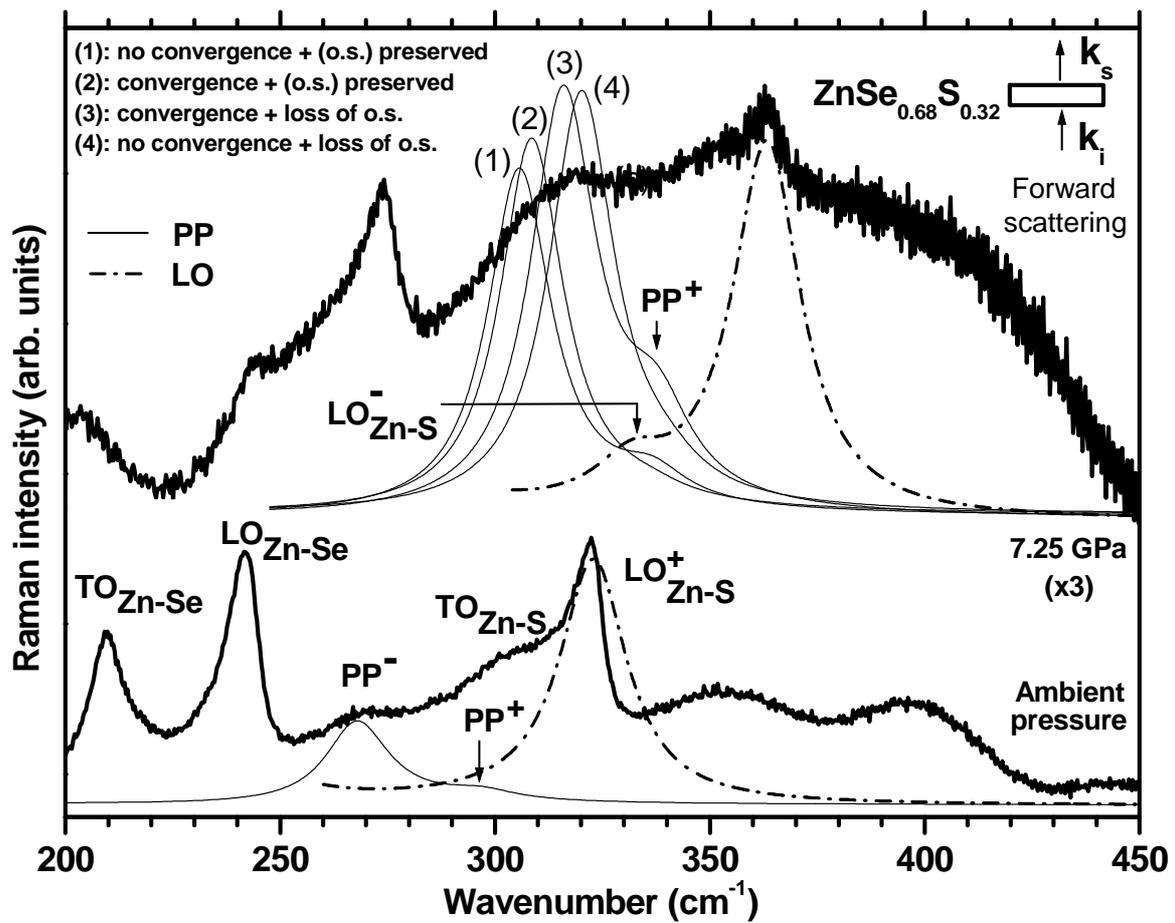

**Figure 4**

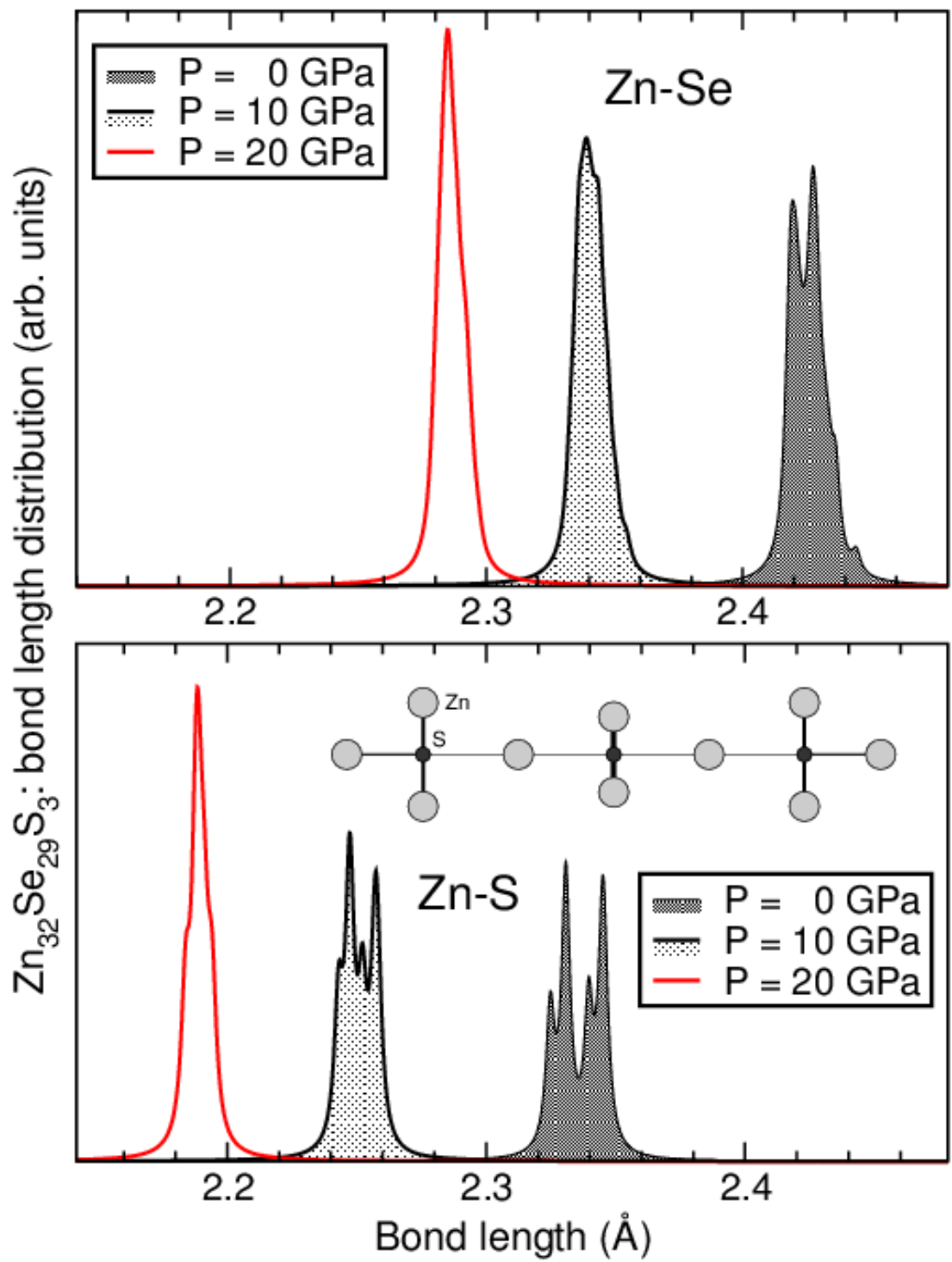

**Figure 5**

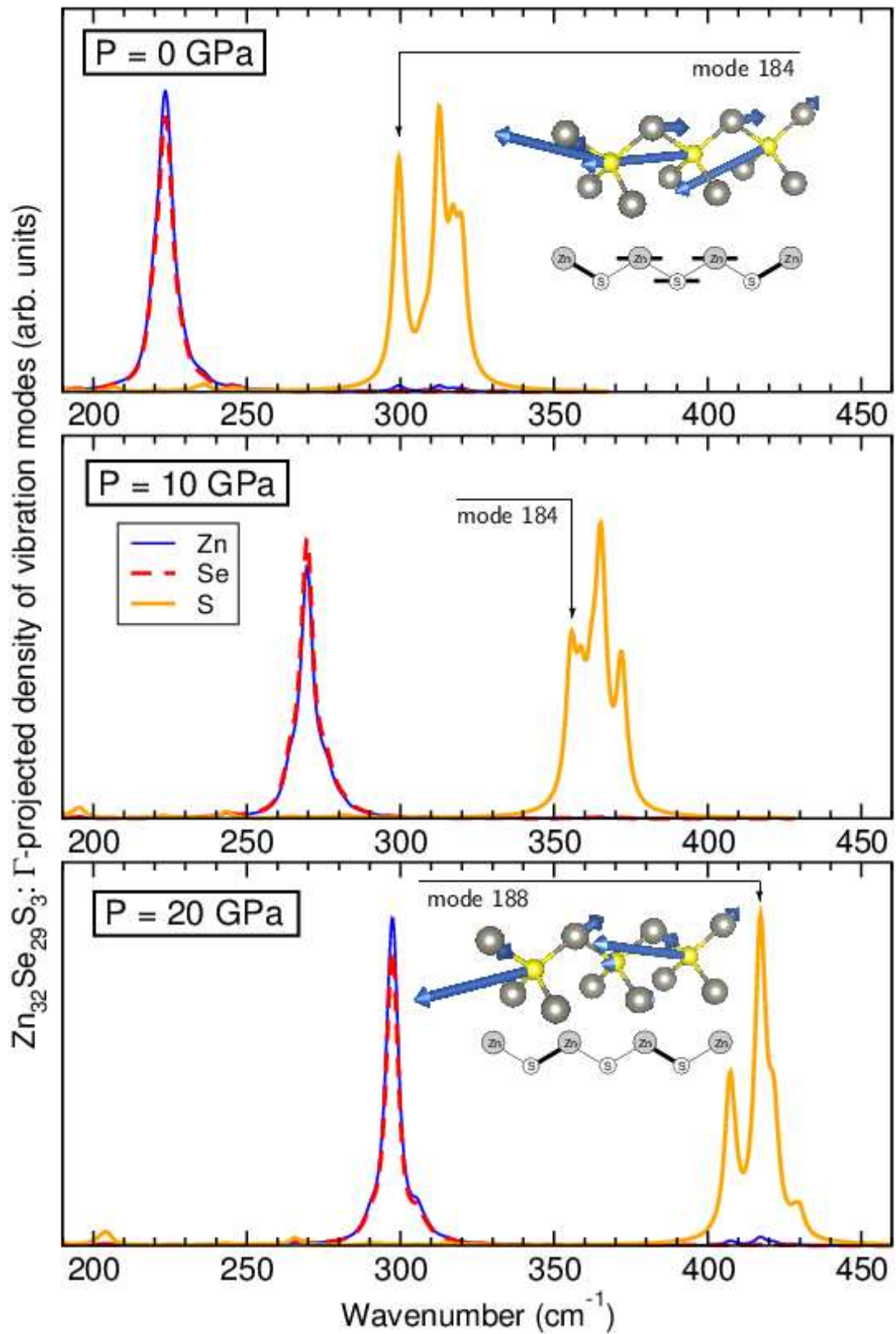

**Figure 6**

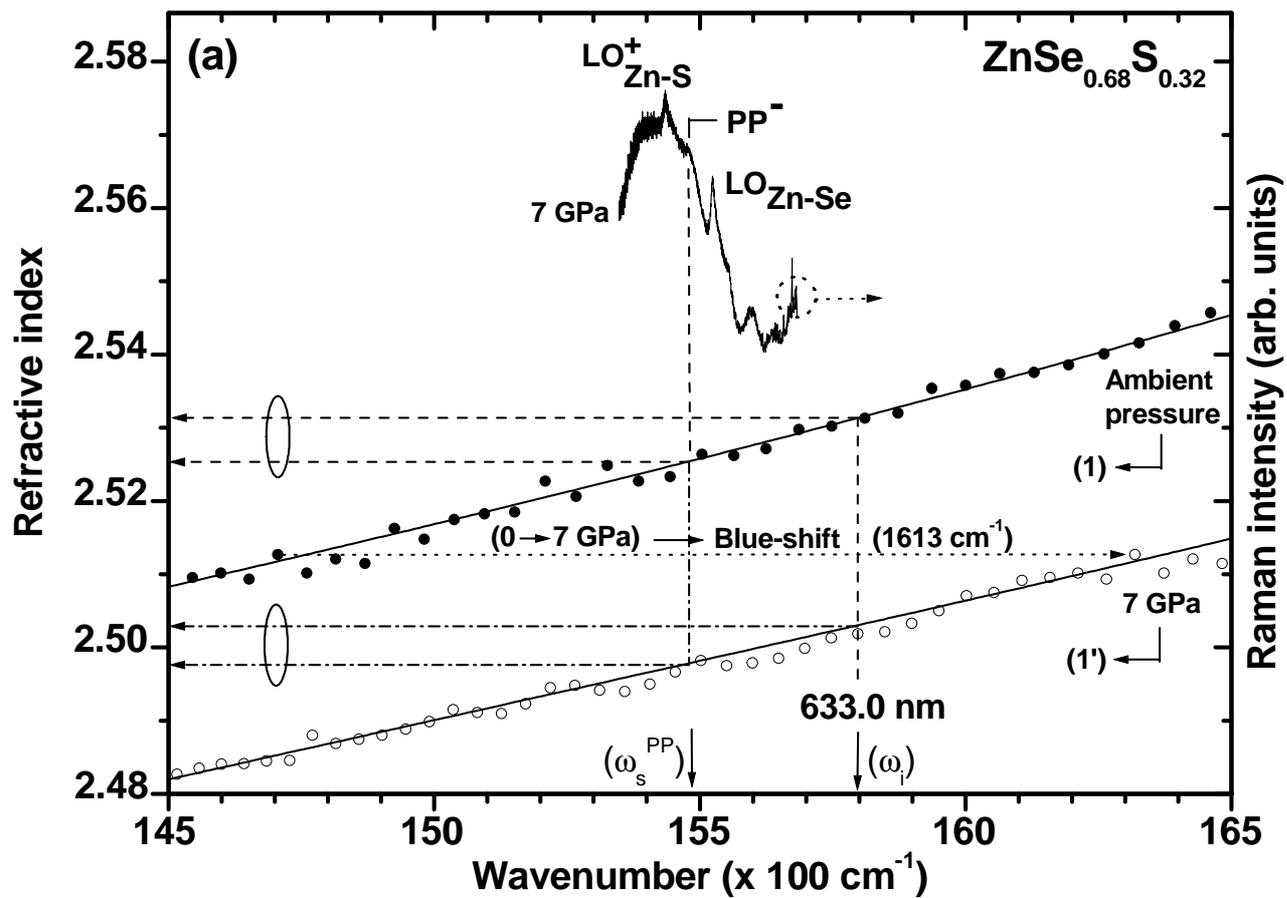

**Figure A1a**

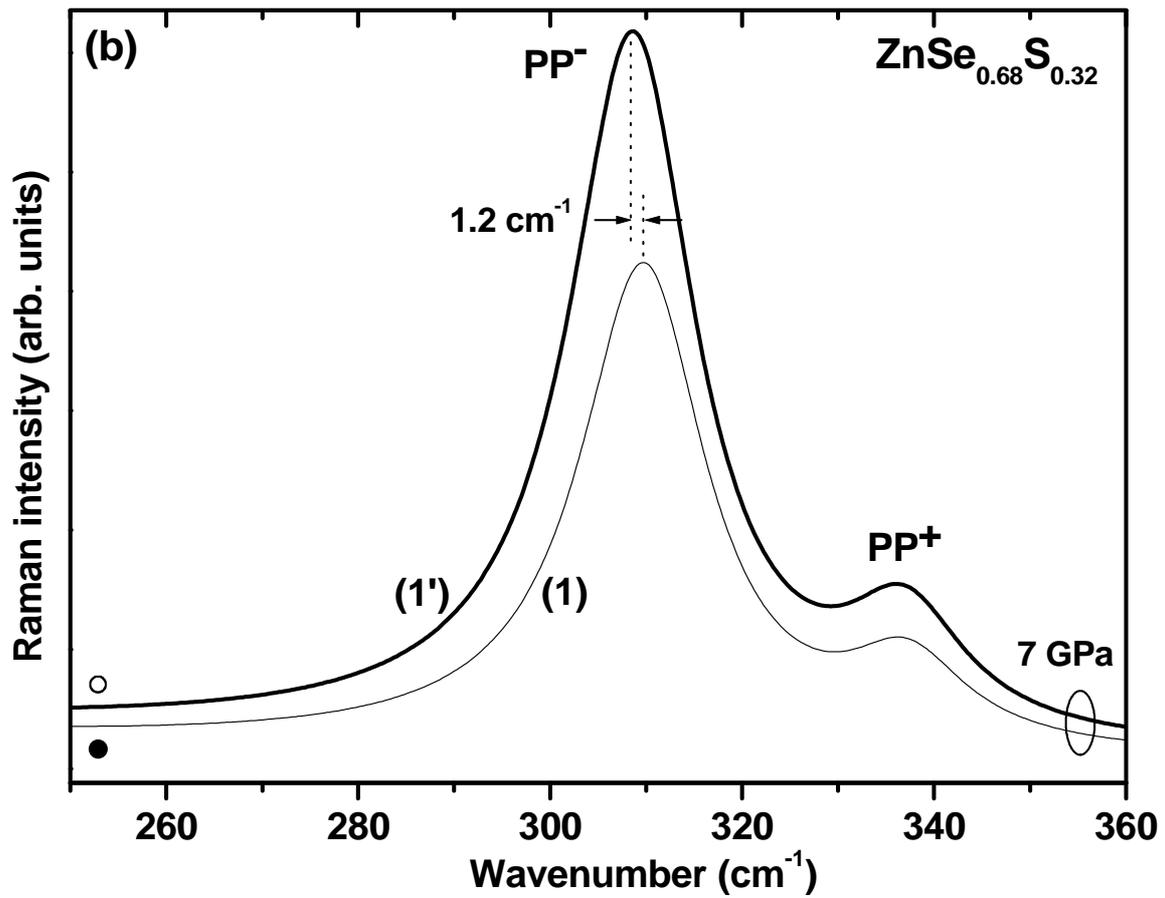

**Figure A1b**